\documentclass[12pt,a4paper,twoside]{article}
\usepackage{epsfig}

\newcommand{\be}{\begin{equation}}
\newcommand{\ee}{\end{equation}}

\newcommand{\PRL}{{\it Phys.~Rev.~Lett.~}}
\newcommand{\JMP}{{\it J.~Math.Phys.~}}
\newcommand{\NP}{{\it Nucl.~Phys.~}}
\newcommand{\PL}{{\it Phys.~Lett.~}}
\newcommand{\PR}{{\it Phys.~Rev.~}}
\expandafter\ifx\csname mathbbm\endcsname\relax

\else

\fi
\begin{document}
\begin{titlepage}
\begin{flushleft}  
       \hfill                      {\tt hep-th/9907064}\\
       \hfill                      UUITP-3/99\\
       \hfill                      June 1999\\
\end{flushleft}
\vspace*{3mm}
\begin{center}
{\LARGE On the virial coefficients of nonabelian anyons
\\}
\vspace*{12mm}
{\large Alexios P. Polychronakos\footnote{E-mail:
poly@teorfys.uu.se} \\
\vspace*{5mm}
{\em Theoretical Physics Department, Uppsala University \\
Box 803, S-751 08  Uppsala, Sweden \/}\\
\vspace*{5mm}
and \\
\vspace*{5mm}
{\em Physics Department, University of Ioannina \\
45110 Ioannina, Greece\/}\\}
\vspace*{15mm}
\end{center}

\begin{abstract}
We study a system of nonabelian anyons in the lowest Landau
level of a strong magnetic field. Using diagrammatic techniques,
we prove that the virial coefficients do not depend on the
statistics parameter. This is true for all representations of all
nonabelian groups for the statistics of the particles and relies
solely on the fact that the effective statistical interaction
is a traceless operator.

\end{abstract}

\end{titlepage}

In two spatial dimensions the group relevant to the quantum 
statistics of particles is the braid group \cite{Wu,GMS1}, rather than the
permutation group. As a result, the possibility for non-standard
statistics exists. A well-studied case is (abelian) anyons \cite{LM,
GMS2,Wi}, 
transforming in a unitary abelian representation of the braid group.
Anyons in the lowest Landau level, in particular, are relevant to the 
quantum Hall effect \cite{Hp,ASW} and constitute realizations 
of ideal exclusion statistics \cite{Hd,VO}.

A natural generalization is nonabelian anyon statistics, 
based on nonabelian representations of the braid group. 
These would be the anyonic analogs of parastatistics \cite{G,MG}.
Just as abelian anyons can be thought of as ordinary
(bosonic or fermionic) particles interacting through a non-dynamical
abelian gauge field, nonabelian anyons can be though of as particles
carrying internal degrees of freedom in some irreducible representation
$R$ of a nonabelian group $SU(n)$ and interacting through an appropriate
non-dynamical nonabelian gauge field. What fixes the statistics, then,
is the group $SU(n)$, the representation $R$ and the coupling strength
$g$ of the internal degrees of freedom to the gauge field. 

A field-theoretic approach to achieving such statistics, in analogy
to the abelian case, is to couple the particles to a nonabelian
gauge field with a Chern-Simons action \cite{DJT}. Such a particle-field
model was proposed by Verlinde \cite{V}. $g$, then, is essentially the 
inverse of the coefficient of the Chern-Simons term and, as such, 
inherits the quantization condition
\be
g = \frac{2}{n} ~,~~~n~{\rm integer}
\ee
This condition does not seem to be crucial for the
purely first-quantized approach and, at any rate, will not play any
role in this paper.

It is of interest to derive the thermodynamics and statistical mechanical
properties of nonabelian anyons in order to probe the possibility of
new physics deriving from the nonabelian nature of the system. In a
recent paper, Isakov, Lozano and Ouvry \cite{ILO} examined these questions
for the simplest case of $SU(2)$ anyons in the fundamental (spin-half)
representation. They found that the virial coefficients up to the
fifth one do not depend on the statistics parameter $g$. They conjectured
that this holds for all the coefficients and posed the question of a possible
underlying symmetry that explains this vanishing dependence. 

The purpose of this note is to give a complete proof of the independence
of all virial coefficient of this model on the statistics parameter $g$,
valid for any group and any representation. It is based on a diagrammatic
expression of the cluster coefficients which is useful in deriving them
in a simple way and reveals their scaling properties with
the volume. It will be apparent that the only feature of the statistical 
interaction which is relevant for this result is that it is a traceless
operator in the space of internal degrees of freedom of the particles.

We repeat here the main results for the system that will be used in
this paper, as presented in \cite{ILO}. The model consists of $N$
non-interacting spinless particles on the plane with internal degrees 
of freedom transforming in some finite-dimensional unitary irreducible
representation
$R$ of $SU(n)$. (We shall refer to these degrees of freedom as 
flavor.) In the gauge where the hamiltonian of the particles is free,
the nontrivial statistics manifests in the wavefunction
of the system, which is not single valued.
Under an exchange of particles following a path belonging to an element 
of the braid group the wavefunction
transforms in some nonabelian representation of the braid group
parametrized by the irrep $R$ of $SU(n)$ and a statistics parameter $g$. 
In principle an abelian part can
also be included, parametrized by a second coupling constant $\alpha$,
endowing the particles with (abelian) anyonic statistics. The contributions
of the abelian and nonabelian part to the statistical mechanis decouple,
however, as will be apparent in the sequel, so we are not going to be
concerned with the abelian part. The particles are taken as bosons as far
as their abelian statistics is concerned.

In analogy with the abelian case, we can perform a singular nonabelian 
gauge transformation that makes
the wavefunction single-valued and bosonic, at the expense of introducing a
gauge field coupling the particles.
We also introduce an external strong constant magnetic field $B=2\omega_c /e$,
as well as an external rotationally invariant harmonic oscillator potential
of frequency $\omega$ (which serves as a `box' to bound the particles).
Upon extracting from the wavefunction an analytic nonabelian factor 
that accounts for its short-distance analytic and braiding behavior 
and a gaussian factor, we are left with an effective hamiltonian reading
\begin{eqnarray}
H &=& \sum_i \left(-2\partial_i {\bar \partial}_i 
-(\omega_t - \omega_c ) z_i \partial_i 
-(\omega_t + \omega_c ) {\bar z}_i {\bar \partial}_i +\omega_t \right)
\nonumber \\
&-&2g \sum_{i<j} T_i^A T_j^A \left( \frac{{\bar \partial}_i -
{\bar \partial}_j}{z_i - z_j} - \frac{\omega_t - \omega_c}{2} \right)
\end{eqnarray}
In the above $z = x + i y$ is a complex coordinate on the plane,
$\partial \equiv \partial/\partial_z$ is the corresponding derivative, 
and $\omega_t^2 = \omega_c^2 + \omega^2$. 
$T_i^A$ are generators of the group $SU(n)$ in the
$R$-representation, each acting in the flavor space of particle $i$; so 
$T^A$ are $d_R \times d_R$ dimensional matrices and $A=1, \dots n^2 -1$.
Summation over repeated indices is always implied.

All homogeneous analytic wavefunctions are eigenstates of
the above hamiltonian. When $B>0$ the analytic wavefunctions become
degenerate in the pure magnetic field limit $\omega \to 0$ and
constitute the lowest Landau level (LLL) for the system. For large
$B$ all higher levels will acquire a large gap and decouple. 
Good analytic behavior of the wavefunction near coincidence points
in that case requires $g>0$. Conversely, for $B<0$ we can extract an
anti-analytic nonabelian factor from the wavefunction and arrive at
an analogous expression for $H$. In that case, it is the
anti-analytic wavefunctions that constitute the LLL and
we must have $g<0$. From now on we will consider the case $B,g>0$,
the opposite one being similar. We shall also assume that $g$ is
not too big, so that no new states descend to the LLL from the
excited spectrum.

The end result is that on states in the LLL the hamiltonian assumes
the form
\be
H = H_o + S
\ee
where $H_o$ is the hamiltonian of a non-interacting bosonic
system and $S$ is the statistics part coupling the internal
degrees of freedom of the particles:
\be
H_o = N \omega_t + (\omega_t - \omega_c) \sum_i z_i \partial_i ~,~~
S = g (\omega_t - \omega_c ) \sum_{i<j} T_i^A T_j^A
\ee
The spectrum of the above hamiltonian can be easily obtained.
$H_o$ essentially counts the degree of homogeneity in $z_i$ of the 
analytic wavefunction, which can then be chosen a homogeneous 
polynomial in $z_i$. $S$ can be expressed as
\be
S = \frac{g}{2} (\omega_t - \omega_c ) \left[ 
\left( \sum_i T_i^A \right)^2 - \sum_i \left( T_i^A \right)^2 \right] 
\ee
Under the total flavor group with generators $T^A = \sum_i T_i^A$,
states transform in the tensor product of $N$ $R$-irreps
$R \times \cdots R$, which can be decomposed into irreducible
components. On states transforming under an irreducible representation
$R_t$ of the total flavor, $S$ becomes
\be
S = \frac{g}{2} (\omega_t - \omega_c ) \left[ C_2 (R_t ) - N C_2 (R) \right]
\ee
where $C_2 (R)$, $C_2 (R_t )$ are the quadratic Casimir of $R$ and $R_t$.
The total wavefunction must carry the $R_t$ representation of total flavor 
and be symmetric under total particle exchange (coordinate and internal 
degrees of freedom). This calls for some group theory for constructing
the states \cite{ILO}. From this spectrum the partition function, 
cluster and virial coefficients can be calculated.

This approach was followed in \cite{ILO} (for $R$ the spin-half 
of $SU(2)$) and the first few virial
coefficients in the thermodynamic limit were thus calculated. We shall 
take here an alternative route, based on a diagrammatic expansion.
The facts central to the derivation are:

\noindent 1. The cluster and virial coefficients at the thermodynamic 
limit can be calculated by taking the strength of the external
potential to zero (corresponding to taking the volume $V$ of the
`box' to infinity). The correct scaling limit for the $k$-th cluster
coefficient is \cite{CGO,O}
\be
\frac{1}{k\beta (\omega_t - \omega_c )} \to V \frac{\omega_c}{\pi}
\ee

\noindent 2. The statistical interaction $S$ is of order $1/V$.

\noindent 3. The statistical interaction $S$ is a sum of two-body terms,
each of which is traceless with respect to the internal space of each
particle.

We now give the rules for the path-integral representation of the
system. (For a more detailed discussion see \cite{P}.)
The $N$-body partition function $Z_N$ can be expressed as a many-body
path integral in periodic euclidean time $\beta$. For short, we shall
call such path-integral configurations diagrams. Since the particles 
are identical, the configuration at time $\tau = \beta$ can be any
permutation of the one at $\tau=0$. This means that the paths of
particles can braid and interchange as they go round the periodic time
direction. Such periodically connected paths of $p$ particles constitute
one `thread' wrapping $p$ times around the time circle. Appropriate
symmetry factors must be included in each diagram to avoid overcounting
of degrees of freedom.

Further, since the particles have color degrees of freedom, each path
is also labeled by a color index $a= 1, \dots d_R$. Summation over all
possible values of such indices is assumed.

The interaction $S$ can be taken into account perturbatively. It is 
two-body and instantaneous, so each insertion corresponds to coupling
two distinct particle paths at a given time. Since it acts on the
flavor space of the two particles, it will change the flavor index on
the two paths before and after the interaction, say from $a$ to $b$ on
one and from $c$ to $d$ on the other. The strength of this interaction
is given by the matrix element
\be
S_{ab;cd} = g (\omega_t - \omega_c ) (T^A )_{ab} (T^A )_{cd}
\ee
The symmetry factors of diagrams with such insertions are modified, 
since the paths connected by $S$ are obviously singled out.
A typical configuration for the path integral in the case of five
particles and two insertions of $S$ is depicted in fig. 1.a, where the
constraints of periodicity for the paths and their flavors have been
taken into account. 

\begin{center}
\leavevmode
\epsfysize=7cm
\epsfbox{dia1.eps}
\end{center}

\noindent For our purposes only the topology and connectivity of these
diagrams will be important, so we will depict them in the simplified
fashion of fig. 1b.
\bigskip

\begin{center}
\leavevmode
\epsfysize=7cm
\epsfbox{dia2.eps}
\end{center}

The grand partition function for the system $\cal Z$ is given by the 
sum of the $N$-body partition functions for all $N$ weighted by fugacity
factors $z^N =e^{\mu \beta N}$ with $\mu$ the chemical potential:
\be
{\cal Z} = \sum_{N=0}^\infty Z_N z^N
\ee
As such, it is the sum of all many-body diagrams. The grand potential
$\Omega$ is the logarithm of $\cal Z$ and, by the standard argument,
it will be given by the sum of all {\it connected} diagrams. 
Two parts of a diagram are disconnected if the particle paths of
each diagram do not mix with the other and if there are no interactions
$S$ coupling the two diagrams. The coefficients $b_k$ of the expansion of
$\Omega$ in powers of $z$ are the cluster coefficients:
\be
\Omega = \sum_{k=1}^\infty b_k z^k
\ee
Therefore, the $k$-th cluster coefficient $b_k$ is simply the sum of 
all connected $k$-particle diagrams.

We come, now, to the question of determining $b_k$ in the thermodynamic
limit. We must isolate, in the class of $k$-body connected diagrams, the
leading contribution in $V$ (or, equivalently, in $(\omega_t - 
\omega_c )^{-1}$), which, for a proper extensive behavior,  must be of
order $V$. To achieve this, note that each topologically connected part
of a diagram, consisting of a single thread looping $p$ times,
in the absence of interaction insertions is of order $V$. Indeed, this
is simply the $p$-th cluster coefficient of noniteracting bosons coming
in $d_R$ flavors, which is properly of order $V$. (Alternatively, if the
infrared regulator were a flat `box' rather than an oscillator potential,
the factor $V$ would come from the translation invariance of the connected
diagram within the box.) 

Thus, if a diagram contains $q$ topologically connected components, it will
be, a priori, of order $V^q$. For it to be connected, there must be
enough insertions of $S$ to connect the $q$ components to each other.
We must have a minimum number of $q-1$ insertions in order to fully
connect the components in a tree-like topology (fig. 2). Since each insertion
$S$ contributes a factor $1/V$, such minimally connected diagrams are
of order $V$. Any further insertion of $S$ will give a sub-leading in
$V$ diagram. In fact, by simple topological counting arguments, we see 
that the number of loops in non-minimally connected diagrams counts the
sub-leading powers of $1/V$.

\begin{center}
\leavevmode
\epsfysize=7cm
\epsfbox{dia3.eps}
\end{center}

So far we concluded that $b_k$ will be given by the sum of all minimally
connected $k$-particle diagrams with any number of components $q$ 
($1\le q \le k$). Now comes the final observation: each tree must have
two or more `endpoints,' that is, components where only one
insertion of $S$ connects. The entire thread of this diagram must
clearly carry the same flavor index $a$; thus, the corresponding
matrix element for the insertion $S$ connecting to this diagram is
$S_{aa;bc}$, with $b,c$ the flavor indices connecting at the other
end of the insertion. Upon summing over $a$ we have
\be
S_{aa;bc} = g (\omega_t - \omega_c ) (T^A )_{aa} (T^A )_{bc}
= g (\omega_t - \omega_c ) (T^A )_{bc}~{\rm tr} T^A = 0
\ee
Therefore, all such diagrams vanish. The only surviving diagram is
the one with a single topologically connected component and no $S$
insertions, which reproduces the cluster coefficient of free bosons
with $d_R$ flavors. Since virial coefficients are uniquely expressed
in terms of cluster coefficients, we have proved that the virial
coefficients of the system are independent of their nonabelian statistics,
for any $R$ of any $SU(n)$.

The above reasoning can also be used to show that the contribution
of the abelian part is the same as in the absence of the nonabelian
part. An abelian part can be included by appending a $U(1)$ generator
$T^0 = Q$ to the $T^A$, proportional to the unit matrix. The trace 
in the above insertion, then, would give
\be
\sum_a S_{aa;bc} = g (\omega_t - \omega_c ) (T^0 )_{aa} (T^0 )_{bc}
= g d_R Q^2 (\omega_t - \omega_c ) \delta_{bc} 
\ee
This is a contribution proportional to an anyonic parameter
$\alpha = g Q^2 d_R$. The effect of the insertion on the flavor
indices of the remaining diagrams is effectively suppressed (since,
due to $\delta_{bc}$, $b=c$). 
Repeating the argument with all endpoint graphs, we eventually
reduce the whole graph to a set of components with non-interacting
flavor indices and the standard abelian statistics interaction
between the components. The cluster coefficients are simply 
$d_R$ times the single-flavor anyonic coefficients.

We remark here that the above techniques can be used to
easily obtain the subleading in $1/V$ contributions to the cluster and
virial coefficients. To each component of the diagram at least two
$S$ insertions are attached (else the diagram vanishes by the previous
argument). Summing over the flavor indices on a component where $m$
insertions attach gives a term proportional to
\be
g^m (\omega_t - \omega_c )^m ~ D_R^{A_1 \dots A_m}
\ee
where the $m$-index symbol $D_R$ is
\be
D_R^{A_1 \dots A_m} = {\rm tr} ( T^{A_1} \dots T^{A_m} )
\ee
and the total diagram involves multiplying the $D$-symbols of each
component and contracting the group indices $A_i$ according to the
connectivity of the components through $S$-insertions. A diagram
with $q$ components and $q+s-1$ insertions will be of subleading
order $1/V^s$ and of order $g^{q+s-1}$ in the statistics parameter.
Since $q$ can range from 1 to $k$ for a $k$-particle diagram we
conclude that the $1/V^s$ correction to the cluster coefficient
$b_k$ will be a polynomial in $g$ with powers ranging from
$g^s$ to $g^{k+s-1}$.

The task of calculating the above corrections simplifies further
in the special case that $R$ is the fundamental of $SU(n)$. In
that case, a well-known completeness relation simplifies $S$:
\be
S_{ab;cd} = g (\omega_t - \omega_c ) \sum_A (T^A )_{ab} (T^A )_{cd}
= g (\omega_t - \omega_c ) \frac{1}{2} (\delta_{ad} \delta_{cb} -
\frac{1}{n} \delta_{ab} \delta_{cd} )
\ee
So $S$ is the sum of a part that simply interchanges the flavor
indices of the strands that it couples plus a part proportional
to the identity operator. The evaluation of diagrams in this case
becomes a simple matter with no group theory required.

Having described the above, we should still point out that the
$1/V$ corrections obtained this way are specific to the `harmonic
box' regularization of the system. They are, thus, likely
not universal and of little interest.

Concluding, the results of this paper are somewhat disappointing,
since they indicate that no new physics are expected at the
thermodynamic limit from any nonabelian statistics of particles
at the LLL. The result seems completely generic since it relies
on little else than the very nonabelian nature of the statistics,
that is, the vanishing of the trace of its generators. Still it is
expected that nonabelian statistics {\it will} influence the
properties of systems {\it not} in the LLL. A calculation of the
properties of such systems along the lines presented here may be
an interesting endeavor.

I am thankful to S.~Isakov and to S.~Ouvry for discussing their
results on nonabelian anyons prior to publication, and to the
Les Houches 1998 organizers for hosting an exciting summer
school where the ideas in this paper were initiated.


\begin{thebibliography}{99}

\bibitem{Wu} Y.S.~Wu, \PRL {\bf 52} (1984) 2103.

\bibitem{GMS1} G.A.~Goldin, R.~Menikoff and D.H.~Sharp, \PRL {\bf 54}
(1985) 603.

\bibitem{LM} J.M.~Leinaas and J.~Myrheim, Nuovo Cimento {\bf 37}
(1977) 1.

\bibitem{GMS2} G.A.~Goldin, R.~Menikoff and D.H.~Sharp, \JMP {\bf 21}
(1980) 650 and {\bf 22} (1981) 1664.

\bibitem{Wi} F.~Wilczek, \PRL {bf 48} (1982) 1144 and {\bf 49} (1982) 957.

\bibitem{Hp} B.I.~Halperin, \PRL {\bf 52} (1984) 1583 and 2390.

\bibitem{ASW} D.~Arovas, J.R.~Schrieffer and F.~Wilczek, \PRL {\bf 53}
(1984) 722.

\bibitem{Hd} F.D.M.~Haldane, \PRL {\bf 67} (1991) 937.

\bibitem{VO} A.~Dasni\`eres de Veigy and S.~Ouvry, \PRL {\bf 72} (1994) 600.

\bibitem{G} H.S.~Green, \PR {\bf 90} (1953) 270..

\bibitem{MG} A.M.L.~Messiah and O.W.~Greenberg, \PR {\bf B136} (1964) 248
and {\bf B138} (1965) 1155.

\bibitem{DJT} S.~Deser, R.~Jackiw and S.~Templeton, \PRL {\bf 48} (1982)
975 and {\it Ann.~Phys.~(NY)} {\bf 140} (1982) 372.

\bibitem{V} E.~Verlinde, `A note on braid statistics and the nonabelian
Aharonov-Bohm effect,' in {\it Modern Quantum Field Theory}, World
Scientific, Singapore 1991.

\bibitem{ILO} S.~Isakov, G.~Lozano and S.~Ouvry, \NP {\bf B552} (1999) 667.

\bibitem{CGO} A.~Comtet, Y.~Georgelin and S.~Ouvry, {\it J.~Phys.}
{\bf A22} (1989) 3917.

\bibitem{O} K.~Olausen, cond-mat/9207005, Trondheim Univ.~prepr.~13 (1992).

\bibitem{P} A.P.~Polychronakos, \PL {\bf B365} (1996) 202 and \NP 
{\bf B474} (1996) 529.

\end{thebibliography}
\end{document}